\documentclass{osa-article}

\newcommand{\Ea}{\ensuremath{{\cal E}_1}}
\newcommand{\Eb}{\ensuremath{{\cal E}_2}}
\newcommand{\Ec}{\ensuremath{{\cal E}_3}}

\newcommand{\Ed}{\ensuremath{{\cal E}_{1,2}}}
\newcommand{\Er}{\ensuremath{{\cal E}_{\rm R}}}
\newcommand{\Eo}{\ensuremath{{\cal E}_{1,2,3}}}

\journal{osajournal}

\articletype{Research Article}

\begin{document}

\title{Dynamics of resonantly excited excitons in MoSe$_2$ and WS$_2$ single-layers monitored with four-wave mixing}

\author{Tomasz Jakubczyk,\authormark{1,2} Miroslav Bartos,\authormark{3,4} Lorenzo Scarpelli,\authormark{5} Karol Nogajewski,\authormark{3,6} Marek Potemski,\authormark{3,6} Wolfgang Langbein,\authormark{7} and Jacek Kasprzak\authormark{1,*}}
\address{\authormark{1}Univ. Grenoble Alpes, CNRS, Grenoble INP, Institut N\'{e}el, 38000 Grenoble, France\\
\authormark{2}Department of Physics, University of Basel, 4056 Basel, Switzerland\\
\authormark{3}Laboratoire National des Champs Magn\'{e}tiques Intenses, CNRS-UGA-UPS-INSA-EMFL, 25 Av. des Martyrs, 38042 Grenoble, France\\
\authormark{4}Central European Institute of Technology, Brno University of Technology, Purkynova 656/123, 61200 Brno, Czech Republic\\
\authormark{5}ARC Centre of Excellence for Engineered Quantum Systems, Macquarie University, Sydney, New South Wales, Australia\\
\authormark{6}Faculty of Physics, University of Warsaw, ul. Pasteura 5,02-093 Warszawa, Poland\\
\authormark{7}School of Physics and Astronomy, Cardiff University, The Parade, Cardiff CF24 3AA, UK}
\email{\authormark{*}jacek.kasprzak@neel.cnrs.fr} 



\begin{abstract}
We investigate dynamics of resonantly excited excitons in single-layers of MoSe$_2$ and WS$_2$ down to 4.5\,K. To this end, we measure the delay dependence of the heterodyne four-wave mixing (FWM) amplitude induced by three, short laser pulses. This signal depends not only on the population of optically active excitons, which affects the absorption of the probe, but also on the population of optically inactive states, by interaction-induced energy shift, influencing the refractive index experienced by the probe. As such, it offers insight into density dynamics of excitons which do not directly couple to photons. Reproducing the coherent signal detected in amplitude and phase, the FWM delay dependence is modeled by a coherent superposition of several exponential decay components, with characteristic time constants from 0.1\,picosecond up to 1\,nanosecond. With increasing excitation intensity and/or temperature, we observe strong interference effects in the FWM field amplitude, resulting in progressively more complex and nonintuitive signal dynamics. We attribute this behaviour to increasingly populated exciton dark states, which change the FWM field phase by the relative effect on absorption and refractive index. We observe that exciton recombination occurs on a significantly longer timescale in WS$_2$ with respect to MoSe$_2$, which is attributed to the dark character of exciton ground state in the former and the bright in the latter.
\end{abstract}

\section{Introduction}
Carrier relaxation dynamics in semiconductor devices, for example in optical amplifiers, is a relevant issue in modern optoelectronics and has been intensively studied using heterodyne pump-probe spectroscopy\,\cite{BorriOC99a, DommersAPL07, CesariJQE09, LangbeinAPL10}. Such methods of ultrafast coherent spectroscopy are also employed in fundamental studies of excitons (EX), bound complexes of electrons and holes, hosted by emerging material systems and their nanostructures, such as colloidal quantum dots\,\cite{MasiaPRL12}, nanoplatelets\,\cite{NaeemPRB15},
perovskites\,\cite{BeckerNanoLett18}, single-layers (SLs) of
semiconducting transition metal dichalcogenides\,\cite{ScarpelliPRB17} (TMDs) and their van der Waals heterostructures\,\cite{JakubczykACSNano19}.

In the SL limit, TMDs become direct band gap materials, with the gap located at K$^{+/-}$ points of the hexagonal Brillouin zone. The optical response, even at room temperature, is dominated by EXs with a binding energy of several hundreds meV. With such a high binding energy --- stemming from the two-dimensional confinement, reduced dielectric screening and high carrier effective masses --- light-matter interaction is expected to be enhanced with respect to conventional semiconductors, like GaAs. This is experimentally confirmed by a strong excitonic absorption in TMD SLs of around 10\%. Also, they accommodate circularly polarized EXs, where the helicity of the transition
($\sigma+\,/\,\sigma-$) is linked with the index $(+/-)$ of the K valley. Thus, apart from spin-allowed (bright) transitions, there are also spin-forbidden (dark) ones, between different valleys. EXs with the in-plane momentum beyond the light cone cannot recombine radiatively either and are also called dark. The multitude of available dark states and their various feeding channels\,\cite{MoodyJOSAB16}, as well as the energetic structure varying from material to material, is expected to strongly affect the EX dynamics in the family of TMDs.

Time-resolved photoluminescence experiments\,\cite{RobertPRB16, GoddePRB16, PlechingerPSS17} performed on TMD SLs have shown bi-exponential or multi-exponential decays on a timescale of a few hundred picosecond. Owing to non-resonant excitation, the modeling
and interpretation of photoluminescence is blurred due to the numerous intermediate states in the scattering pathway from the initially excited electron-hole pairs to the emission of the bright EX states. Conversely, resonant driving with a short laser pulse, directly imprints a well controlled density of bright EXs within the light cone into a well defined valley (K$^{+}$, K$^{-}$ or both).

This can be conveniently accomplished by performing four-wave mixing (FWM) spectroscopy\,\cite{LangbeinOL06}. The latter was recently employed to investigate dynamics of resonantly driven EX population in MoSe$_2$ SLs\,\cite{ScarpelliPRB17}. Considering the excitonic landscape in TMD SLs, these initial FWM studies have established several dominating processes in the EX density dynamics, like direct radiative recombination competing with scattering and evolution of density in dark state reservoirs (out of the light cone
direct spin-allowed, direct spin-forbidden, indirect spin-allowed, and indirect spin-forbidden). These experiments were however performed only down to 77\,K and measurements at lower temperatures down to 4.2\,K have not been carried out. Recently, an example of EX population dynamics measured via FWM was reported and discussed in Ref.\,[\citenum{JakubczykACSNano19}] using MoS$_2$ heterostructure.

In this work, we study dynamics of resonantly excited EXs in bare SLs of MoSe$_2$ and WS$_2$ down to 4.5\,K. We present a comprehensive set of FWM measurements on a timescale from 0.1-1300\,ps, combining excitation power ($P$) and temperature ($T$) dependencies. Increasing $P$ and/or $T$, the dark states become increasingly more populated, resulting in more and more complex FWM delay traces. In these experiments, we observe prominent interference effects in the FWM amplitude, revealing different scattering channels between several EX reservoirs. We also conclude that exciton relaxation occurs on a significantly longer timescale in WS$_2$ with respect to MoSe$_2$, which is tentatively attributed to dark character of the exciton ground state in WS$_2$ and the bright in MoSe$_2$. Following Ref.\,[\citenum{ScarpelliPRB17}], the FWM amplitude \emph{versus} delay is modeled by a coherent superposition of up to five components of complex exponential decays, generating highly non-intuitive delay traces, yet reproducing the ensemble of our measurements very well.

\section{Experiment}

\subsection{Samples}
\begin{figure}
   \begin{center}
   \begin{tabular}{c}
   \includegraphics[height=9cm]{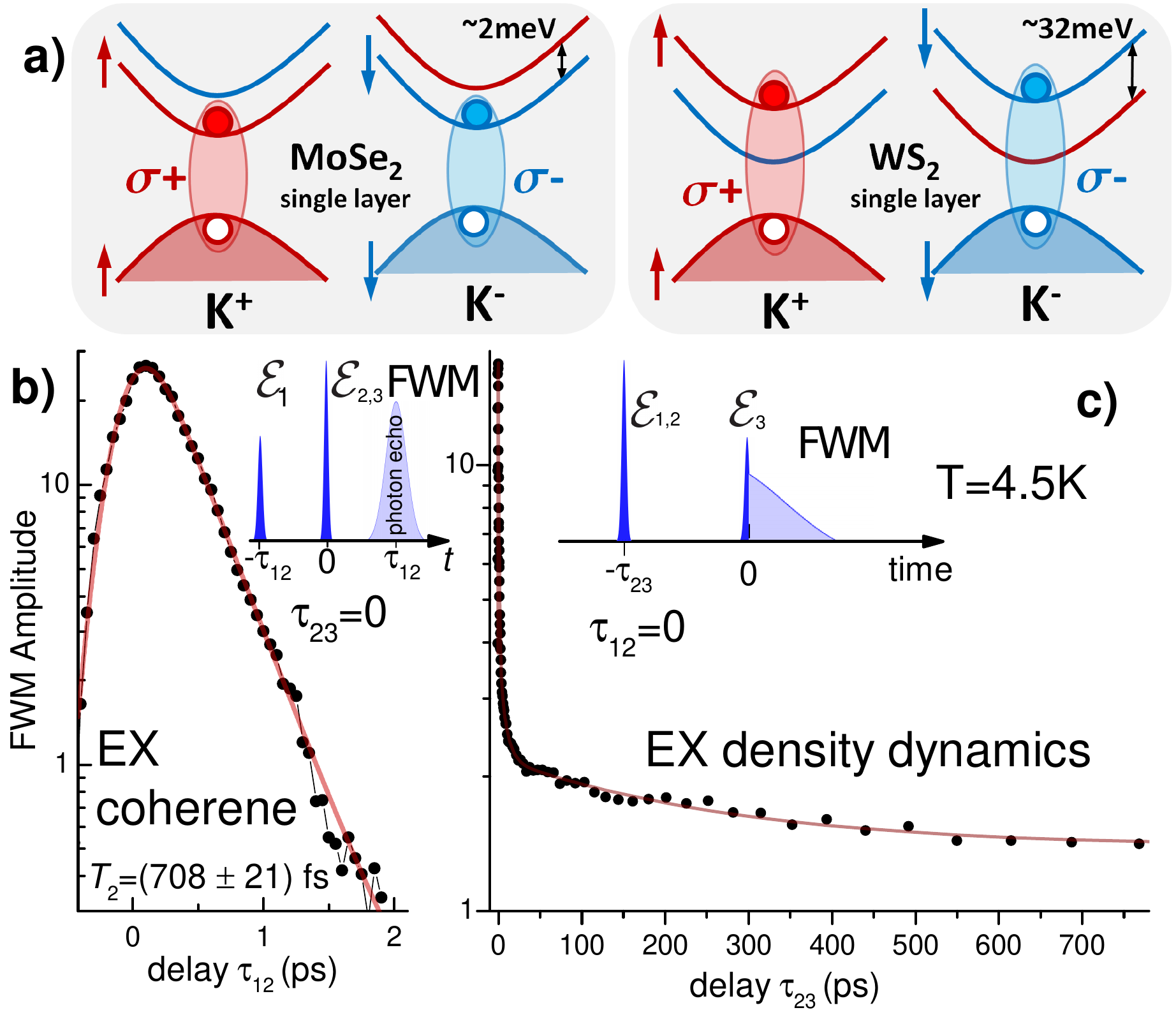}
   \end{tabular}
   \end{center}
   \caption[example]
   { \label{fig:fig1} Direct exciton transitions in single-layers of transition metal dichalcogenides and their observables probed with four-wave mixing spectroscopy. a)\,Schema of band structure of TMD SLs, showing the highest lying valence band and the lowest lying pair of conduction bands, at the K$^{+}$ and K$^{-}$ points. Owing to the reverse energy splitting, in MoSe$_2$ (WS$_2$) the transition of the lowest energy is optically (in)active, bright (dark). b,\,c)\,Representative measurements on MoSe$_2$ SL at T=4.5K. b)\,Time-integrated FWM amplitude as a function of delay $\tau_{12}$, for $\tau_{23}$=0, as depicted in the inset, reflecting EX coherence dynamics. c)\,Time-integrated FWM amplitude as a function of delay $\tau_{23}$, for $\tau_{12}=0$, as depicted in the inset, reflecting EX population dynamics.}
\end{figure}

We here use two kinds of TMD SLs: MoSe$_2$ and WS$_2$, the same as the ones as explored in Ref.\,[\citenum{JakubczykNanoLett16}] and
Ref.\,[\citenum{Jakubczyk2DMat17}], respectively. In both cases, the SLs were produced by means of polydimethylsiloxane-based exfoliation of bulk crystals, purchased from HQ Graphene, and transferred onto a Si substrate with a 90\,nm SiO$_2$ surface layer and then inserted into an optical He-flow cryostat. In MoSe$_2$, the lowest conduction band and the highest valence band have the same spin, therefore the EX with the lowest energy is bright (optically active). It is worth to note that, the next conduction band hosting an electron with the opposite spin, and thus contributing to the dark EX, lies only 1.5\,meV away\,\cite{Lu19}, around twenty times closer than previous
theoretical predictions\,\cite{DeryPRB15}. In WS$_2$ SLs though the configuration is inverse, \emph{i.e.}, the lowest conduction band and the highest valence band have opposite spin, therefore the lowest lying EX is dark. The next conduction band, sharing the same spin projection as the valence band, is around 30\,meV higher, producing the spin-allowed bright exciton. The corresponding energy diagrams for both materials are depicted in Fig.\,\ref{fig:fig1}\,a.

\subsection{Why four-wave mixing microscopy\,?}

SL-TMDs explored in these experiments have been fabricated as bare flakes deposited directly on SiO$_2$/Si substrates. They are in direct contact with the environment, thus are prone to various microscopic and macroscopic disorder factors, such as: strain, wrinkling, flake deformations and cracks, lattice defects and vacancies, variation of charge state and of dielectric constant in the substrate. Their optical response is therefore affected by spectral inhomogeneous
broadening\,\cite{MoodyNatCom15, JakubczykNanoLett16}, $\sigma$, which varies across the flake\,\cite{Jakubczyk2DMat17} from nearly zero to a few tens of meV. When examining optical properties of TMDs it is desirable to use a spectroscopic tool capable to separate homogeneous and inhomogeneous contributions within the EX's spectral lineshape. In order to avoid adding up responses from areas exhibiting different optical properties, one would also preferentially employ microscopy, where diffraction limited excitation beams restrict spatial averaging to sub-micron areas.

The heterodyne FWM technique, exploited in this work, complies with these two requirements. FWM is a nonlinear polarization in the material, proportional to $\mu^4\Ea^{\star}\Eb\Ec$, where $\mu$ is the transition oscillator strength and $\Eo$ are three femto-second laser pulses. To study EXs in MoSe$_2$ SL at a light wavelength around 750\,nm, we use a Ti:Sapphire laser as pulse source (Tsunami-Femto, from Spectra-Physics). Conversely, for EXs in WS$_2$ SL at around 590\,nm, we use optical parametric oscillator (Inspire, provided by Radiantis). We use a different set of optics adapted for these two wavelengths. The temporal chirp at the sample was pre-compensated using a grating-based pulseshaper to achieve Fourier-limited pulses.

FWM is here detected through optical heterodyning, enabling co-linear propagation of $\Eo$. This permits microscopy configuration, which has been required to measure coherent nonlinear response of individual excitons\,\cite{KasprzakJOSAB12, KasprzakNJP13, MermillodPRL16}. First, $\Eo$ are modulated by distinct radio frequencies $\Omega_{1,\,2,\,3}$ around 80\,MHz using acousto-optic deflectors. $\Eo$ propagate co-linearly and are focussed onto the sample down to the diffraction limit using a microscope objective of NA=0.65. In the detection path, the reflected signal is interfered with the reference beam $\Er$ modulated at the phase-locked linear combination of radio-frequencies $\Omega_3+\Omega_2-\Omega_1$, picking up the FWM response from the reflection. Then, background-free interference between $\Er$ and FWM is recorded as non-oscillating spectral interference on a CCD camera, which is installed at the output of the imaging spectrometer\,\cite{LangbeinOL06}. FWM amplitude and phase are obtained \emph{via} spectral interferometry. Owing to the limited phase stability and lack of phase referencing\,\cite{DelmontePRB17} in the current experiments, only the FWM amplitude is exploited.

As depicted in Fig.\,\ref{fig:fig1}\,b, in a presence of $\sigma$, FWM forms a photon echo centered at $t=\tau_{12}$, where the latter is the delay between $\Ea$ and $\Eb$. The time-integrated (TI) FWM amplitude as a function of $\tau_{12}$ measures the EX coherence dynamics, which in the simple case decays exponentially. From the decay constant, one retrieves the microscopic dephasing time $T_2$, which can be translated to the full-with at half-maximum of the homogeneous broadening $\gamma=2\hbar/T_2$. An example of such a measurement performed on MoSe$_2$ at T=4.5\,K, yielding $T_2\simeq0.7\,$ps, is presented in
Fig.\,\ref{fig:fig1}\,b.

There are two principal channels causing EX dephasing: radiative
decay (occuring in sub-ps range) and pure dephasing. The latter is
due to the EX-EX scattering, such that the phase imprinted onto the
EX cloud by $\Eo$ is rapidly lost. As a result the EX coherence
withstands only a few ps. Instead, a part of EX density transferred
to the dark states is present and evolves on a much longer
timescale; up to at least 1\,ns, as presented in
Fig.\,\ref{fig:fig1}\,c. To measure EX density dynamics we set
$\tau_{12}=0$. $\Ea$ and $\Eb$ induce EX density $\Ea^{\star}\Eb$,
slowly oscillating at the frequency $\Omega_2-\Omega_1$. The EX
density in different states affects via EX-EX interaction the EX
polarization created by \Ec, inducing the detected FWM. The
amplitude and phase of the FWM signal is therefore a probe of the
EX-EX interaction between the optically probed transition and the EX
states containing the density. To rephrase, after the delay
$\tau_{23}$, the last pulse $\Ec$ arrives and triggers the FWM,
which is proportional to the remaining EX density weighted by their
mutual interactions. The dynamics of this combined effect is probed
by reading out the TI-FWM amplitude \emph{versus} $\tau_{23}$, as
depicted in the inset of Fig.\,\ref{fig:fig1}\,c. The figure also
shows that, at low $T$ and weak $P$, more than 90\% of the EX
population is gone already after only a few ps, owing to the
radiative decay and simultaneous scattering to the EX dark states.

Another possible channel leading to the loss of density is the EX
diffusion: during $\tau_{23}$, EXs can be ejected out of the
sub-$\mu$m excitation spot\,\cite{PereaNanoLett19}. We have looked
for such propagation by performing spatially-resolved FWM
experiments. Namely, by spatially displacing (typically, by several
$\mu$m) $\Ed$ with respect to the probe beam $\Ec$ while detecting
FWM, one becomes sensitive to the EX in-plane propagation effects.
We have performed numerous experiments in such configurations,
providing no evidence for EX diffusion in our samples. In fact,
owing to the inhomogeneous broadening in these bare SLs, EXs are
expected to be spatially localized and thus exhibiting nearly no
diffusion at low temperatures.

The following of this work is devoted to presentation and modeling of such $\tau_{23}$-dependences measured at MoSe$_2$ and WS$_2$ for different $P$ and $T$.

\subsection{Modeling exciton density dynamics measured \emph{via} heterodyne four-wave mixing.}
To describe the loss of EX density when increasing delay
$\tau_{23}$, the proposed fitting function $R$ is a multi-exponential decay for $\tau_{23}>0$ and zero otherwise:
$R(\tau_{23})\propto \sum_{i=1}^{n}A_i\theta(\tau_{23})\exp{(\imath\phi_i-\frac{\tau_{23}}{\tau_i})}$.
Because we employ a coherent detection scheme, it is sensitive to the phase of each of the decay processes. Therefore, $R$ is a complex function, providing characteristic times, amplitudes and phases of $n$ processes that are involved, $(\tau_i,\,A_i,\,\phi_i)$.

To mimic the experiments, the temporal form of the pulse train, employed to drive FWM, must be taken into account. Firstly, to take into account the finite pulse duration, $R$ is convoluted with a Gaussian characterized by a FWHM of $2\sqrt{\ln(2)}\tau_0$ given by the laser autocorrelation. Secondly, owing to the lase repetition period $T_R\simeq13\,$ns, the signal might not fade away completely before the arrival of the following pulse sequence, but the leftover proportional to $\exp({-\frac{T_R}{\tau_i}})$ is present. When adding up such contribution from the second pulse sequence onward
over an infinite number of pulses in the heterodyne detection
scheme, the pile-up term is obtained as the geometric sum
$\exp({-\frac{T_R}{\tau_i}})+\exp({-\frac{2T_R}{\tau_i}})+\exp({-\frac{3T_R}{\tau_i}})+\exp({-\frac{4T_R}{\tau_i}})+...=\frac{\exp({-\frac{T_R}{\tau_i}})}{1-\exp{(-\frac{T_R}{\tau_i}})}$.
$R$ therefore takes the form:

\begin{align}
R(\tau_{23})=\sum_{i=1}^{n}A_i\{[\exp({\frac{T_R}{\tau_i}})-1]^{-1}+\frac{1}{2}[(1+{\rm
erf}(\frac{\tau_{23}}{\tau_0}-\frac{\tau_{0}}{2\tau_i})]\}\times\exp{(\imath\phi_i-\frac{\tau_{23}}{\tau_i}+\frac{\tau_0^2}{4\tau_i^2})}
\label{Eq:ComplexFit}
\end{align}

Finally, to fit the FWM amplitude we take absolute value of
$R(\tau_{23})$, namely $|R(\tau_{23})|$, yielding a rather lengthy, yet still analytical form.

\section{RESULTS}
\subsection{MoSe$_2$ single layer}
\begin{figure}
   \begin{center}
   \begin{tabular}{c}
   \includegraphics[height=9cm]{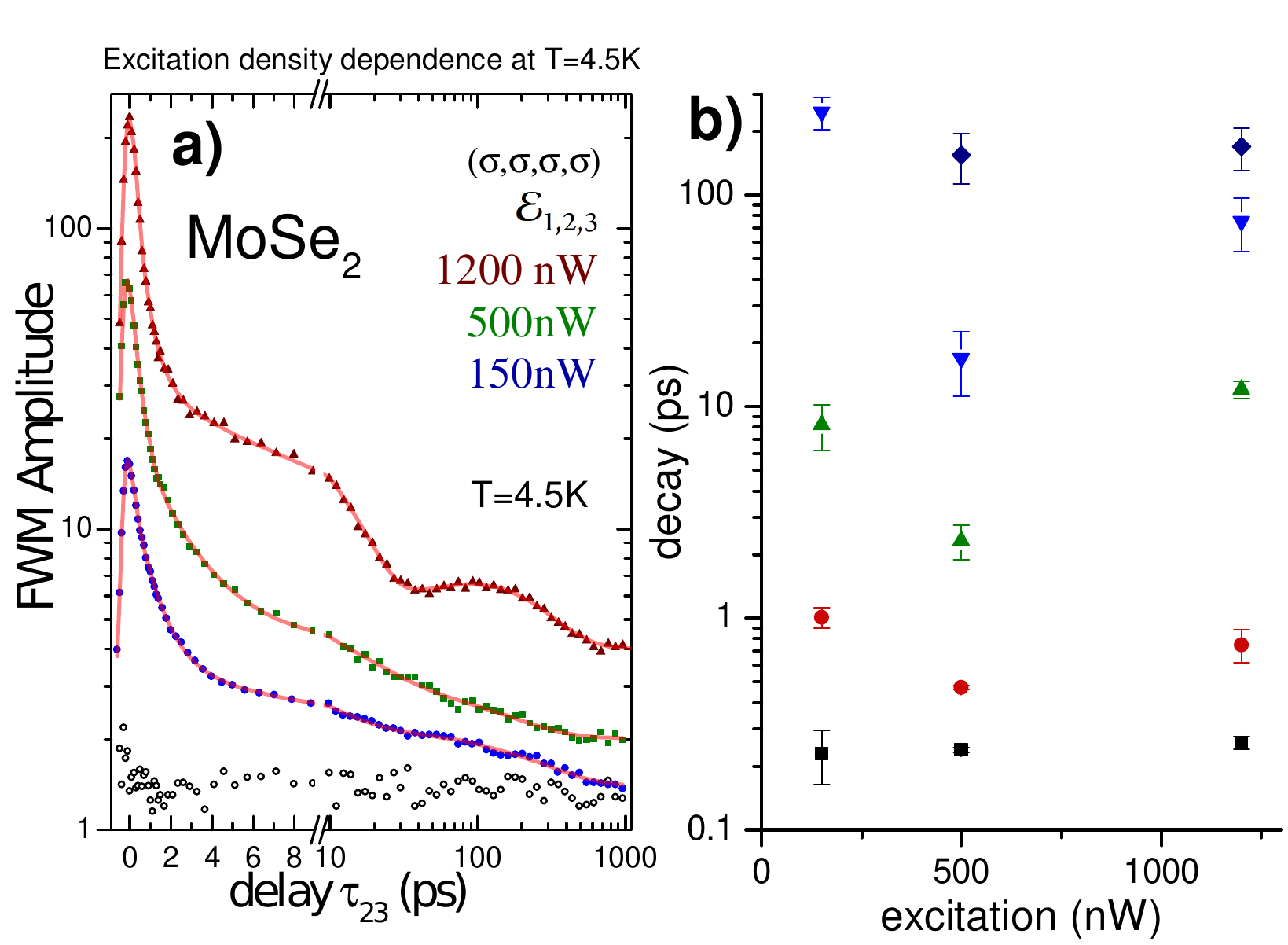}
   \end{tabular}
   \end{center}
   \caption[example]
   {\label{fig:fig2} Exciton density dynamics in MoSe$_2$ SL measured with four-wave mixing. Excitation power $P$ as indicated at $T=4.5\,$K. For the highest $P=1200\,$nW an oscillation observed between 10\,ps and 1000\,ps is due to the interference between different decay components, enabled by increased amplitude of the ones featuring longer decay constants. In all figures, fits are given with semi-transparent traces.}
   \end{figure}

In Fig.\,\ref{fig:fig2}\,a we present the EX density dynamics measured on MoSe$_2$ SL at $T=4.5\,K$ for three different driving powers $P$, as indicated. We estimate that exciting with 1\,$\mu$W generates around $10^{11}\,$EX/cm$^2$. Qualitatively, for the ensemble of our data we distinguish: i)\,signal at $\tau_{23}<0$ owing to the pile-up effect, ii)\,when approaching $\tau_{23}=0$ from negative values, we see a rise on a 100\,fs scale, due to the finite duration of the laser pulses, iii)\,strong, initial decay, by around an order of magnitude in the FWM amplitude over a few hundreds fs scale, attributed to the radiative decay and EX scattering to dark EX states. iv)\,followed after $\tau_{23}>1\,$ps by a much longer decay, due to the population of the dark EX states and their subsequent relaxation and interaction with the bright ones, \emph{i.e.}, EXs within the light cone, obeying spin selection rules for radiative recombination.

The dynamics is fitted with $|R(\tau_{23})|$ yielding several decay constants from 0.2\,ps to 200\,ps, as reported in Fig.\,\ref{fig:fig2}\,b. It is worth to point out that at the highest excitation power, after a dominating initial decay, for $\tau_{23}>10\,$ps a strong modulation is measured and reproduced by the model. This is due to the interference between signals of different decay processes, whose phases are given by the relative effect on absorption and refractive index dynamics. We expect that the phases are largely modified with increasing excitation powers, due to a larger population of dark states reservoir, consistently with our data.

\begin{figure}
   \begin{center}
   \begin{tabular}{c}
   \includegraphics[height=9.5cm]{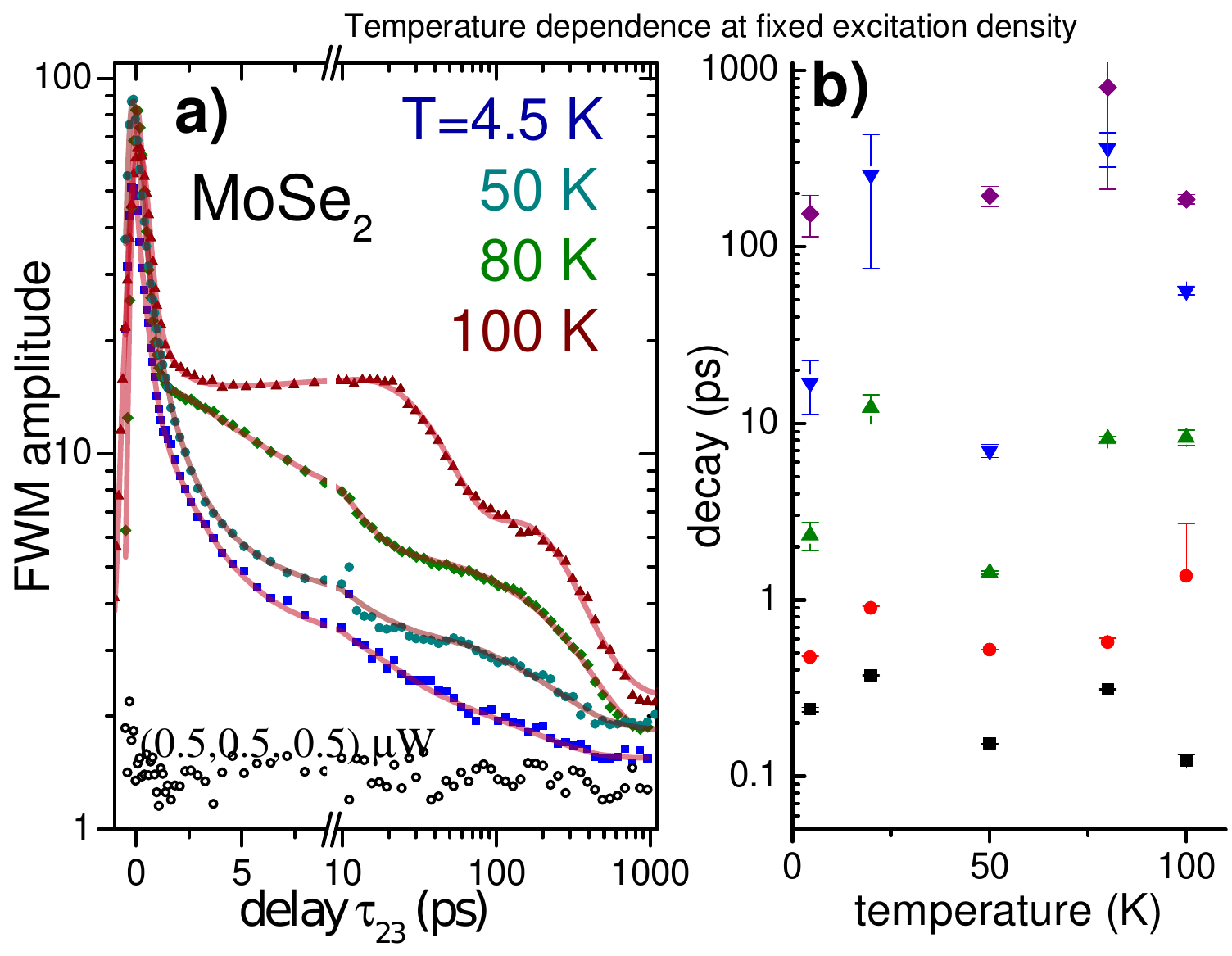}
   \end{tabular}
   \end{center}
   \caption[example]
   { \label{fig:fig3}
Exciton density dynamics in MoSe$_2$ SL measured with four-wave mixing. Temperature dependence, as indicated, at fixed excitation power $P=0.5\,\mu$W. With increasing temperature unusual shapes of the delay traces are measured and reproduced by the modeling. In the fit we observe that the interference sets in when the components showing a longer decay increase their relative amplitudes.}
   \end{figure}

To study the influence of phonon scattering and of thermal distribution across the states, we measure the EX density dynamics for increasing $T$ at fixed $P=0.5\,\mu$W, as shown in Fig.\,\ref{fig:fig3}\,a. Note that for each temperature the pulse wavelength was adjusted to match the EX transition, which was monitored with the reflectance of $\Ea$. Also, the temperature induced spatial drifts were corrected, such that the same sample spot was probed. The dynamics are fitted with $|R(\tau_{23})|$ involving five decay components, explicitly given in Fig.\,\ref{fig:fig3}\,b. For the fastest decay, we observe a decrease from around 300\,fs down to 100\,fs, when increasing $T$ from 5\,K to 100\,K. While no clear trends are observed for other decay components, the interplay of their relative amplitudes induces important changes in the FWM amplitude for $\tau_{23}>2\,$ps. For
the lowest temperatures a monotonous decay of the FWM is observed when increasing $\tau_{23}$, as exemplified for $T=4.5\,$K. From $T=50\,$K onward, a bump starts to build up at $\tau_{23}\simeq100\,$ps. With increasing $T$, the bump progressively shifts to shorter $\tau_{23}$. At $T=100\,$K, after the initial decay, the FWM amplitude increases, reaches maximum at $\tau_{23}=20\,$ps, to finally decay generating another bump at $\tau_{23}=200\,$ps. Again, such a surprising form of the $\tau_{23}$-delay dependence is due to the interference between different decay processes, building up when increasing $T$ owing to increasingly larger amplitude of processes exhibiting longer decay. This is a strong indication of increased population in a multiplicity of dark EX reservoirs with increasing temperature, mainly enabled by increased phonon scattering at higher temperatures.

\subsection{WS$_2$ single layer}
\begin{figure}
   \begin{center}
   \begin{tabular}{c}
   \includegraphics[height=9cm]{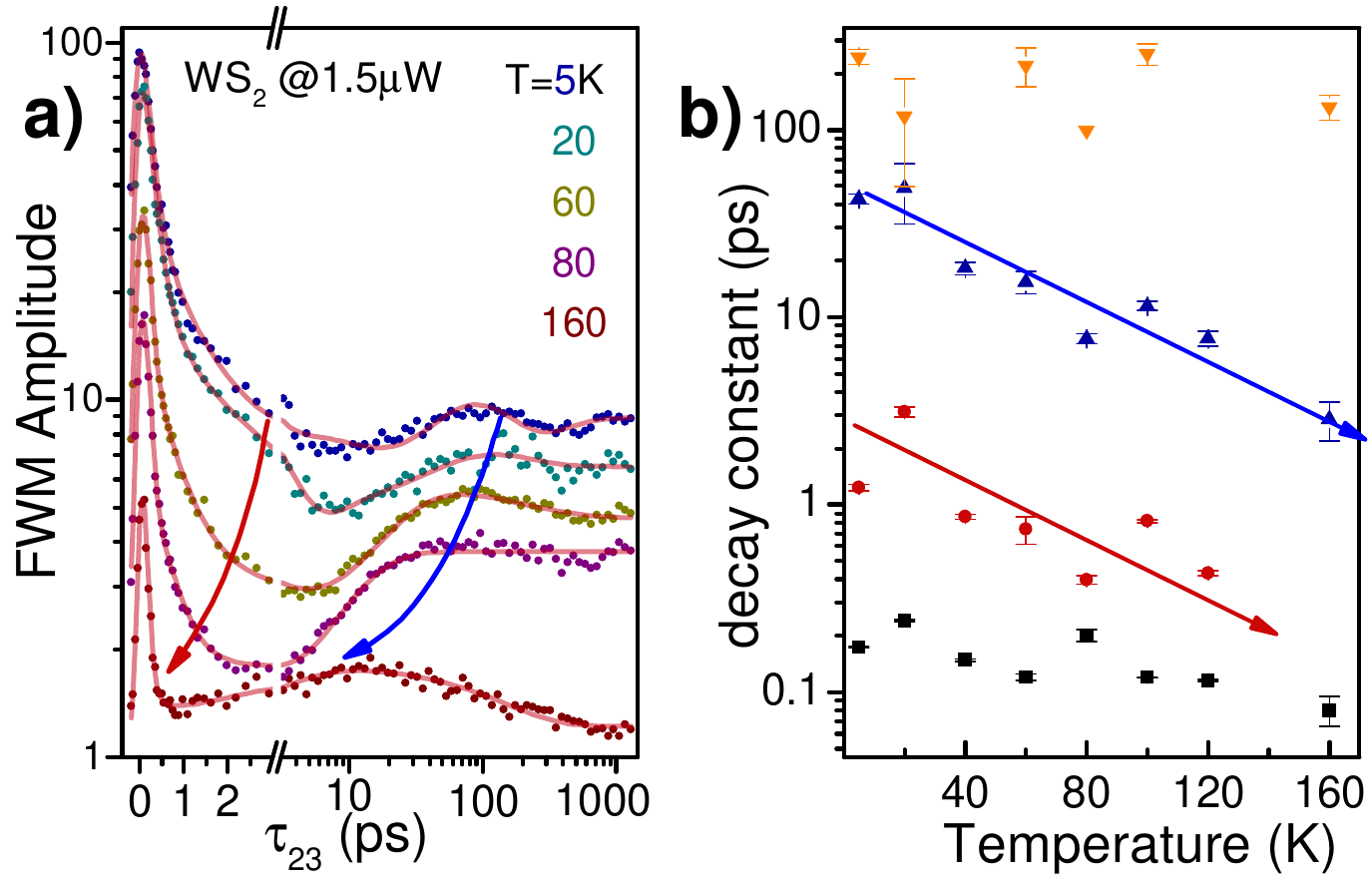}
   \end{tabular}
   \end{center}
   \caption[example]
   { \label{fig:fig4} Exciton density dynamics in WS$_2$ SL measured with four-wave mixing. Temperature dependence, as indicated, at fixed excitation power $P=1.5\,\mu$W. The modeling indicates shortening of the second and the third decay constants with increasing temperature. It is visible in the data as a shift of the bumps towards shorter delays $\tau_{23}$ when increasing the temperature.}
   \end{figure}

We now discuss complementary results obtained on a WS$_2$ SL. In Fig.\,\ref{fig:fig4}\,a we present the EX density dynamics measured at fixed $P=1.5\,\mu$W for different temperatures. At low temperatures a strong pile up effect is observed. Namely, at $T=5\,$K after a dominating initial decay within $0<\tau_{23}<1\,$ps, a quasi-stationary signal with a minor modulation at around 100\,ps is observed, around an order of magnitude above the noise level. Interestingly, with increasing temperature, the initial loss of the signal becomes more pronounced, followed by the recovery, which here is most pronounced for $T=80\,$K. At higher temperatures the recovery at longer $\tau_{23}$ is suppressed, such that a broad bump is visible at $T=160\,$K. Qualitatively, we observe shifting of modulations to shorter delays with increasing $T$, as indicated with red and blue arrows.

The data are again modeled with the $|R(\tau_{23})|$ function containing four components. The fitted decay constants are plotted in Fig.\,\ref{fig:fig4}\,b, from which we conclude that with increasing $T$ the dynamics strongly accelerates: when increasing $T$, values of $\tau_1$, $\tau_2$ and $\tau_3$ decrease. To interpret the data we remind that the bright EX in WS$_2$ is located around $32\,$meV above the optically inactive (spin-forbidden) one. The initial decay corresponds to the radiative decay and the scattering to dark states, also toward the optically inactive EX. The lower the temperature, the more efficiently EX population gets trapped in this
state. In particular, at $T=5\,$K the Boltzmann factor
exp(-32meV/0.43meV)=exp(-74.4) entirely isolates the population from the thermal activation toward the bright EX. Alternative scattering channels are required to bring the population back to the bright EX state. Such a long relaxation time explains a very strong pile up effect observed at low temperatures. With increasing $T$ the thermal scattering from the spin-forbidden to spin-allowed EX state becomes relevant, for example at $T=160\,$K the Boltzmann factor rises to $\exp(-32{\rm meV}/13.7{\rm meV})\simeq0.1$, enabling faster relaxation processes. This results in a faster decay and negligible pile-up at high temperatures, consistent with our findings.

\begin{figure}
   \begin{center}
   \begin{tabular}{c}
   \includegraphics[height=13.5cm]{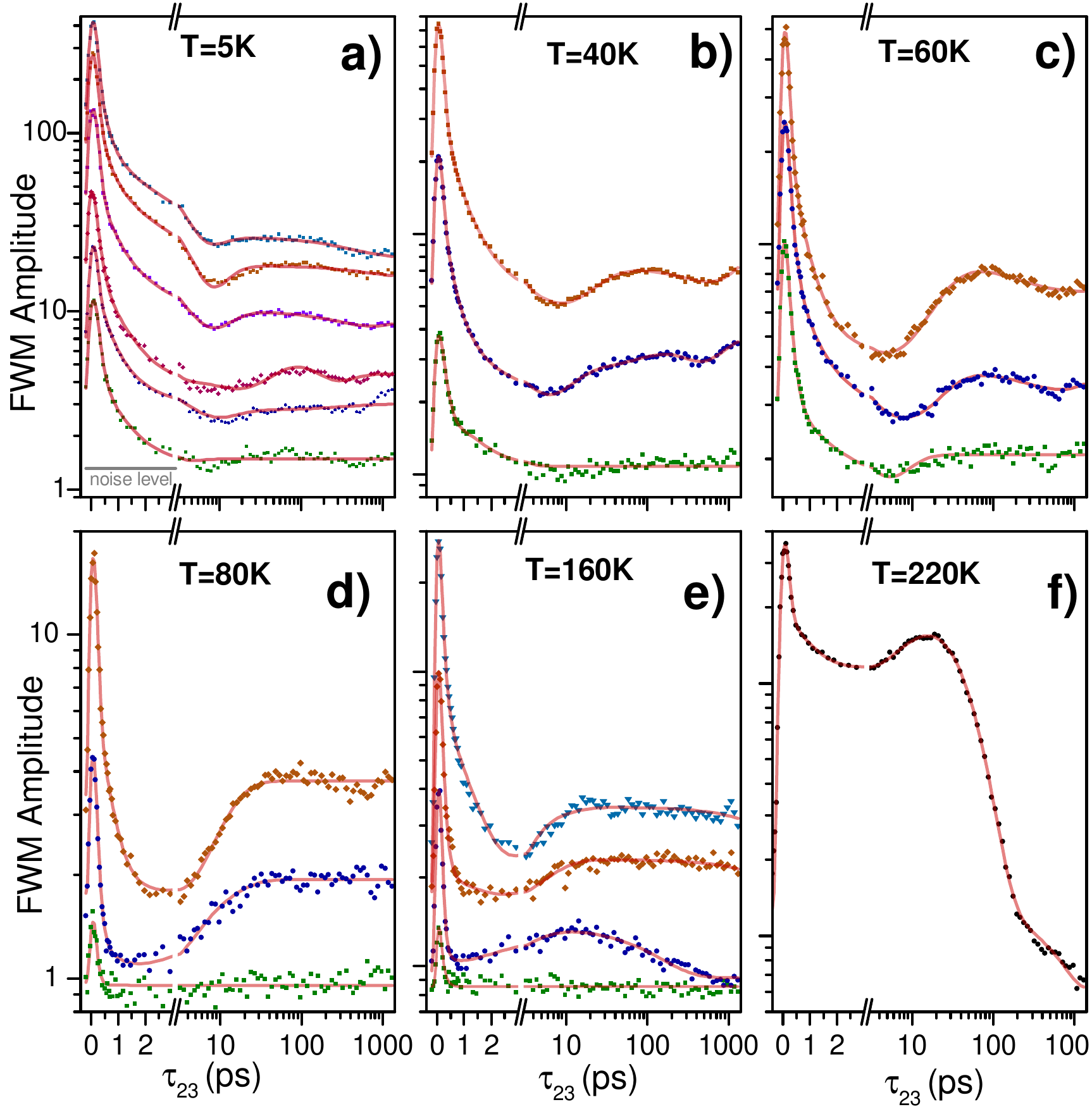}
   \end{tabular}
   \end{center}
   \caption[example]
   { \label{fig:fig5}
Exciton density dynamics in WS$_2$ SL measured with four-wave
mixing. Power dependence measured at different temperatures as
indicated. The used excitation powers are the following:
a)$\,(0.2,\,0.4,\,0.8,\,2.7,\,5.2)\,\mu$W,
b)$\,(0.08,\,0.4,\,1.6)\,\mu$W, c)$\,(0.08,\,0.4,\,1.5)\,\mu$W,
d)$\,(0.08,\,0.4,\,1.5)\,\mu$W, e)$\,(0.4,\,1.5,\,3.6,\,7.5)\,\mu$W,
f)$\,6\,\mu$W.}
   \end{figure}

Finally, we note that even at elevated temperatures, accumulation of the EX density for long $\tau_{23}$ and the pile up effect can be restored by increasing $P$ and thus occupancy of the dark EX states. This can be concluded from Fig.\,\ref{fig:fig5}, showing intensity dependence of the FWM delay traces for different temperatures.

\section{CONCLUSIONS}
To conclude, using heterodyne FWM micro-spectroscopy, we measured temporal dynamics of EXs from 0.1\,ps to 1.3\,ns in MoSe$_2$ and WS$_2$ SLs after resonant excitation with a short laser pulse. The rich data set, including excitation intensity and temperature dependence for both materials, is consistently modeled with a coherent superposition of several exponential decays. They are attributed to direct radiative decay, EX scattering out of the light cone, and spin-flip processes between spin-allowed and spin-forbidden EXs. Owing to the complex character of the data, unconventional interference effects arise in the FWM delay dependence, when different EX reservoirs display comparable occupancy. This permits to gain direct insight into interaction of bright excitons with the dark ones. As an outlook, we propose to use this approach to measure EX radiative lifetimes and evolution of EX dark states in van der Walls heterostructures with engineered thicknesses of encapsulating layers\,\cite{FangPRL19}.
\paragraph{Disclosures} The authors declare no conflicts of interests.

\end{document}